\documentclass[a4paper,UKenglish,cleveref,autoref,thm-restate,numberwithinsect]{lipics-v2021}

\nolinenumbers

\usepackage{amsmath,amssymb,amsfonts}
\usepackage{multirow}
\usepackage[most]{tcolorbox}

\newcommand{\calF}{\mathcal{F}}

\newcommand{\mypara}[1]{\smallskip\noindent\textbf{#1.}}

\newcommand{\temph}[1]{\emph{#1}}

\newcommand{\ia}{\textit{i}}
\newcommand{\ib}{\textit{ii}}
\newcommand{\ic}{\textit{iii}}

\usepackage{enumitem}
\setlist{nosep, leftmargin=*}

\title{Characterising Global Platforms: Centralised, Decentralised, Federated, and Grassroots}
\titlerunning{Characterising Global Platforms}

\author{Ehud Shapiro}{London School of Economics and Weizmann Institute of Science}{}{}{}

\authorrunning{Shapiro}

\Copyright{Ehud Shapiro}

\hideLIPIcs
\ArticleNo{0}



\begin{document}

\maketitle

\begin{abstract}
Global digital platforms are distributed systems designed to serve entire populations, with some already serving billions of people. Here we propose atomic transactions-based multiagent transition systems and protocols as a formal framework to study them; introduce essential agents---minimal sets of agents the removal of which makes communication impossible; and show that the cardinality of essential agents partitions all global platforms into four classes: 
\begin{enumerate}
    \item Centralised (Facebook) -- one (the server)
    \item Decentralised (Bitcoin) -- finite  $>1$ (bootstrap nodes)
    \item Federated (Mastodon) -- infinite but not universal (all servers)  
    \item Grassroots (Scuttlebutt) -- universal (all agents but one)
\end{enumerate}
Our illustrative formal example is a global social network, for which we provide centralised, decentralised, federated, and grassroots specifications via multiagent atomic transactions,  
and prove they all satisfy the same basic correctness properties, yet have different sets of essential agents as expected.
We discuss informally additional global platforms---currencies, ``sharing economy'' apps, AI, and more.

While this may be the first formal characterisation of centralised, decentralised, and federated global platforms, grassroots platforms have two previous definitions.  We prove that (1) the two definitions are incomparable; (2) each implies that all agents but one are essential; and (3) both implications are strict, presenting two platforms with all agents but one essential that are not grassroots under either definition.

This work provides the first mathematical framework for classifying any global platform---existing or imagined---by providing a multiagent atomic-transactions specification of it and determining the cardinality of the minimal set of essential agents in the ensuing multiagent protocol.  It thus
provides a unifying mathematical approach for the study of global digital platforms, perhaps the most important class of distributed systems today. 
\end{abstract}

\noindent This paper is not eligible for the best student paper award.

\newpage
\setcounter{page}{1}

\begin{table}[t]
\centering
\begin{tabular}{|l|c|c|c|}
\hline
\textbf{Class} &  \textbf{Canonical Example} &  \textbf{Essential Agents} &  \textbf{Cardinality}\\
\hline
\textbf{Centralised} & Facebook & The server &  One \\
\hline
\textbf{Decentralised} & Bitcoin~\cite{nakamoto2008bitcoin} & Bootstrap nodes & Finite, $>1$ \\
\hline
\textbf{Federated} & Mastodon~\cite{mastodon} & All servers/all clients & Infinite, not universal \\
\hline
\textbf{Grassroots} & Scuttlebutt~\cite{tarr2019secure} & All agents but one  & Universal \\
\hline
\end{tabular}
\caption{Classes of global platforms by cardinality of essential agents}
\label{table:properties}
\end{table}

\section{Introduction}\label{section:intro}

Global digital platforms are software systems that aim to serve entire populations---all of humanity in some, entire nations in others.  Global platforms have emerged as humanity's primary infrastructure for social interaction, economic exchange, and information flow. Despite their ubiquity and profound influence on billions of lives, we lack a mathematical framework to study their fundamental architectures.  The informal classification of platforms as centralised, decentralised, federated, or peer-to-peer is well-known, but hitherto has been imprecise: the boundaries between classes are vague, the classification is not exhaustive, and it is unclear whether the classes are mutually exclusive.

\mypara{Our Framework}
We propose atomic transactions-based multiagent transition systems and protocols~\cite{shapiro2025atomic} as a formal framework to study and characterize global platforms. In this framework, platforms are modelled as sets of agents whose interactions are governed by atomic transactions---indivisible state changes that can involve multiple participants. 

To demonstrate the viability of this framework for specifying platforms of all four classes, we use a Facebook/X-like social network---conceived as centralised platforms---as our running example, providing atomic-transactions specifications of centralised, decentralised, federated, and grassroots social networks with feeds and followers. 

\mypara{Essential Agents: A Lens for Classification}
We then introduce \emph{essential agents}---minimal sets of agents the removal of which makes communication impossible (only unary transitions may occur). This concept provides a mathematical lens for understanding platform dependencies. The cardinality of essential agents in multiagent protocols provides the first formal characterisation of global platforms, partitioning them into four classes, as shown in Table~\ref{table:properties}.

\begin{table}[t]
\centering
\small
\begin{tabular}{|l|l|l|}
\hline
\textbf{Class} & \textbf{Control} &\textbf{Examples} \\
\hline
\textbf{Centralised} & \textbf{Corporate} &Google Search (5Bn), YouTube (2.5Bn), Gmail (2.5Bn), \\
&&Facebook (3Bn), Instagram (3Bn), WhatsApp (3Bn), Messenger (1Bn), \\
&& TikTok (1.6Bn), Douyin (750M), iPhone (1.5Bn), WeChat (1.4Bn), \\
&& X/Twitter (500M), Discord (200M)~\cite{discord}, Web servers, Database servers \\
\hline
\textbf{Decentralised} & \textbf{Capital} & Bitcoin (\$2.3Trn)~\cite{nakamoto2008bitcoin}, Ethereum (\$500Bn)~\cite{buterin2014next,wood2014ethereum}, \\
&&Tether (\$180Bn), BNB (\$160Bn), Solana (\$110Bn), \\
&& (IPFS~\cite{benet2014ipfs}, DHT/Kademlia~\cite{maymounkov2002kademlia}, BitTorrent~\cite{cohen2003incentives}) \\
\hline
\textbf{Federated} &  \textbf{Local Operators} & Mastodon (15M)~\cite{raman2019challenges}, Matrix (80M)~\cite{matrix2019}, \\
&& ActivityPub networks, Email (SMTP servers) \\
\hline
\textbf{Grassroots} & \textbf{Participants} & 
Scuttlebutt~\cite{tarr2019secure} \\
&& Grassroots social networks$^*$~\cite{shapiro2023gsn}, \\
& &  Grassroots coins and bonds$^*$~\cite{shapiro2024gc,lewis2023grassroots,shapiro2026bonds}, \\
&& Grassroots federations$^*$~\cite{shapiro2025GF}, \\
&& Grassroots Logic Programs$^*$~\cite{shapiro2025glp}, \\
&& $^*$ Mathematically specified, yet to be implemented \\
\hline
\end{tabular}
\caption{Classification of global platforms, with user counts and market valuations where available.}
\label{table:examples}
\end{table}

\mypara{Global Platforms in Practice}
We review the ``common knowledge'' on global platforms of the four classes, exposing their fundamental importance and influence.  See also Table~\ref{table:examples}.\footnote{All numbers in the paper are approximate, as of November 2025; compiled from public reporting and industry trackers and cross-verified by AI.} 

\begin{enumerate}[leftmargin=*]
\item \textbf{Centralised platforms} employ the centralised client-server architecture, where corporations control cloud servers that mediate all interactions among people who use them. Five companies command platforms with over one billion users each: Google/Alphabet (5Bn users; \$3Trn market cap), Meta (3.8Bn; \$2Trn), ByteDance (2Bn; \$350Bn), Apple (1.5Bn; \$4Trn), and Tencent (1.4Bn; \$750Bn).
Collectively, these companies have over \$10Trn market value and reach over 5Bn people. Their governance is \emph{autocratic}: one entity holds all decision-making power, resulting in a handful of people controlling together a substantial portion of the digital lives of more than half of humanity. Facebook, Twitter, Uber, and traditional web services exemplify this architecture, where a server with a designated identity (e.g. facebook.com) remains permanently essential for platform operation. This architecture enables what may be characterized as \emph{corporate control}~\cite{zuboff2019age,zuboff2022surveillance}.

\item \textbf{Decentralised platforms} include global cryptocurrency platforms that employ the blockchain architecture, where distributed networks of nodes maintain replicated ledgers through consensus protocols while users access services through wallets and decentralised applications. Bitcoin~\cite{nakamoto2008bitcoin} (\$2.3Trn market cap) and Ethereum~\cite{buterin2014next,wood2014ethereum} (\$500Bn) exemplify this architecture. The top five platforms collectively exceed \$3Trn in market value.
While envisioned as decentralised alternatives to centralised financial systems, in practice cryptocurrencies exhibit extreme concentration of both ownership and control: Regarding ownership, less than 5\% of cryptocurrency holders, which is less than 0.25\% of the human population, own more than 90\% of their value, exacerbating an already-unprecedented concentration of wealth. Regarding control, it is not only \emph{plutocratic}, but also far from being decentralised: In Bitcoin the top two mining pools control ~55\% of hashrate, and in Ethereum the top four operators control ~50\% of staked ETH. In both mechanisms, control is realized via capital---computational resources (PoW) or staked assets (PoS). We may term this \emph{capital control}~\cite{vitalikplutocracy}.  Decentralised platforms also include IPFS~\cite{benet2014ipfs} and Kademlia-based distributed hash tables~\cite{maymounkov2002kademlia}, which have a finite number of boot nodes that new participants must reach to join the network.  BitTorrent~\cite{cohen2003incentives} is decentralised: its original version relies on a tracker and its trackerless version on DHT boot nodes, so in both, joining a swarm requires reaching a finite set of coordinating agents.

\item \textbf{Federated platforms} employ a distributed server architecture where multiple independent servers interoperate through shared protocols, with users attached to their chosen home server. Examples include Mastodon~\cite{raman2019challenges} (15M users across ~10,000 servers), Matrix~\cite{matrix2019} (80M users), and the broader ActivityPub ecosystem.
While being more distributed than centralised platforms, federated systems maintain a fundamental architectural distinction between servers and clients. Users cannot function without their home server, and server operators exercise control over their domains---deciding moderation policies, user access, and federation relationships. This may be characterized as \emph{control by local operators}.

\item \textbf{Grassroots platforms} aim to provide an \emph{egalitarian, cooperative and democratic} alternative to centralised/autocratic and decentralised/plutocratic global platforms~\cite{shapiro2023grassrootsBA,shapiro2025atomic}.
Unlike the global platforms that may only have a single instance (one Facebook, one Bitcoin), grassroots platforms enable any agent to initiate an instance, have multiple instances operate independently, and have instances coalesce, possibly (but not necessarily) resulting in a single global instance. 
Grassroots platforms were specified for social networks~\cite{shapiro2023gsn}, coins and bonds~\cite{shapiro2024gc,lewis2023grassroots,shapiro2026bonds} and democratic federations~\cite{shapiro2025GF,lewis2023grassroots}. Existing grassroots platforms include 
Scuttlebutt~\cite{tarr2019secure}
\end{enumerate}

\mypara{Contributions}
Our introduction of essential agents as a mathematically-founded notion provides two fundamental contributions to the study of global platforms. 
The cardinality of minimal sets of essential agents yields a comprehensive and mutually exclusive characterisation of all four known platform classes---centralised (one), decentralised (finite, $>1$), federated (infinite, but not universal), and grassroots (universal). This is the first formal characterisation of centralized, decentralized, and federated global platforms, now unified with grassroots computing within a single mathematical framework. 
Grassroots platforms have two previous definitions.  We prove that (1) the two definitions are incomparable; (2) each implies that all agents but one are essential; and (3) both implications are strict, presenting two platforms with all agents but one essential that are not grassroots under either definition.

\mypara{What the formalism reveals}
The framework exposes properties of global platforms that are not apparent from informal descriptions.  First, it shows that social networks do not require consensus: the grassroots social network specification satisfies the same correctness properties as the centralised, decentralised, and federated specifications, yet involves no consensus mechanism, exposing a fundamental architectural mismatch in blockchain-based social networks.  Second, it makes precise the distinction between ``decentralised'' and ``grassroots,'' which are often conflated informally: the cardinality gap between finite and universal essential agents is sharp.

\mypara{Paper organization} Section~\ref{section:framework} presents the mathematical framework: atomic transactions, multiagent transition systems, protocols, liveness, essential agents, and the four platform classes. Section~\ref{section:specifications} uses social networks as a running example, providing transaction-based specifications for all four platform classes, stating correctness and cardinality of essential agents for each. Section~\ref{section:grassroots} presents two formal definitions of grassroots protocols, proves them incomparable, and proves that both imply all agents but one are essential and that both implications are strict. Section~\ref{section:related} reviews related work and Section~\ref{section:conclusion} concludes. Proofs of correctness and cardinality for the social network specifications appear in Appendix~\ref{appendix:proofs}, and Appendix~\ref{section:additional} reviews additional global platforms of all classes.

\section{Mathematical Background}\label{section:framework}

Here, we recall multiagent transition systems~\cite{shapiro2021multiagent}, the notion of grassroots protocols and platforms~\cite{shapiro2023grassrootsBA}, and their specification via multiagent atomic transactions~\cite{shapiro2025atomic}, and introduce the novel notions of transaction equivalence, liveness, and essential agents, and use the cardinality of essential agents to classify global platforms.

\subsection{Atomic Transactions}

We assume a potentially infinite set of agents $\Pi$, but consider only finite subsets of it, so when referring to a particular set of agents $P \subset \Pi$ we assume $P$ to be nonempty and finite.  We use $\subset$ to denote the strict subset relation and $\subseteq$ when equality is also possible.

In the context of multiagent transition systems it is common to refer to the state of the system as \emph{configuration}, so as not to confuse it with the \emph{local states} of the agents. As standard, we use $S^P$ to denote the set of all total functions from $P$ to $S$, and if $c \in S^P$ is a configuration over $S$ and $P$, we use $c_p$ to denote the member of $c$ indexed by $p \in P$, i.e.\ $c_p = c(p)$.

\begin{definition}[Local States, Configuration, Transaction]\label{definition:at}
Given agents $Q \subset \Pi$ and an arbitrary set $S$ of \emph{local states}, a \emph{configuration} over $Q$ and $S$ is a member of $C := S^Q$. An \emph{atomic transaction} over $Q$ and $S$ is a pair of configurations $t = c \rightarrow c' \in C^2$ such that $c \ne c'$, with $t_p := c_p \rightarrow c'_p$ for any $p \in Q$.
\end{definition}
An agent $p$ is an \emph{active participant} in $t$ if $c_p \ne c'_p$. A transaction is \emph{unary} if it has exactly one participant, \emph{binary} if it has two, and \emph{$k$-ary} in general.

\subsection{Multiagent Transition Systems and Protocols}

\begin{definition}[Transition System]\label{definition:ts}
A \temph{transition system} is a tuple $TS = (C, c0, T)$ where $C$ is an arbitrary set of \temph{configurations}, $c0 \in C$ is a designated \temph{initial configuration}, and $T \subseteq C \times C$ is a \temph{transition relation}, with $(c,c') \in T$ also written as $c \rightarrow c' \in T$.
\end{definition}
A transition $c \rightarrow c' \in T$ is \temph{enabled} from configuration $c$. A configuration $c$ is \temph{terminal} if no transitions are enabled from $c$. A \temph{computation} is a (finite or infinite) sequence of configurations where for each two consecutive configurations $(c,c')$ in the sequence, $c \rightarrow c' \in T$. A \temph{run} is a computation starting from $c0$, which is \temph{complete} if it is infinite or ends in a terminal configuration.  A \temph{prefix} of a computation is a finite initial segment of it.

Given agents $P \subset \Pi$, local states $S$ with initial state $s0 \in S$, we denote configurations as $C := S^P$ and the initial configuration as $c0 := \{s0\}^P$.

A transaction and a transition are structurally identical---both are pairs of configurations---but differ in their role.  A transaction is specified over its participants $Q$, the agents whose states are preconditions for the transaction to occur, and says nothing about agents outside $Q$; different transactions may have different sets of participants.  A transition, by contrast, is over a fixed set of agents $P$, as required for the construction of a transition system.  Given a set of transactions, each over its own set of participants, the closure operator (Definition~\ref{definition:closure}) induces from them a set of transitions over a fixed $P$, in which non-participants remain stationary.

\begin{definition}[Transaction Closure]\label{definition:closure}
Let $P\subset \Pi$, $S$ a set of local states, and $C:=S^P$.  We say $p$ is \emph{stationary} in a transition or transaction $t = c \rightarrow c'$ if $c_p = c'_p$.
For a transaction $t=(c\rightarrow c')$ over local states $S$ with participants $Q$, the \temph{$P$-closure of $t$}, $t{\uparrow}P$,  is the set of transitions over $P$ and $S$ defined by:
$$
t{\uparrow}P := \begin{cases} \{ t' \in C^2  :
\forall q\in Q.(t_q = t'_q) \wedge \forall p\in P\setminus Q.(p\text{ is stationary in }t')\} & \text{if } Q\subseteq P \\
\emptyset & \text{otherwise}
\end{cases}
$$
If $R$ is a set of transactions, each $t\in R$ over some $Q$ and $S$, then the 
\temph{$P$-closure of $R$}, $R{\uparrow}P$, is the set of transitions over $P$ and $S$ defined by:
$$
R{\uparrow}P := \bigcup_{t\in R} t{\uparrow}P
$$
\end{definition}
Namely, the closure over $P\supseteq Q$ of a transaction $t$ over $Q$ includes all transitions $t'$ over $P$ in which members of $Q$ do the same in $t$ and in $t'$, and the rest remain in their current (arbitrary) state.

A set of transactions $R$ over $S$, each with participants $Q\subseteq P$, defines a multiagent transition system over $S$ and $P$ as follows:

\begin{definition}[Transactions-Based Multiagent Transition System]\label{definition:tbmats}
Given agents $P \subset \Pi$, local states $S$ with initial local state $s0\in S$, 
and a set of transactions $R$, each $t\in R$ over some $Q\subseteq P$ and $S$, the \emph{transactions-based multiagent transition system} over $P$, $S$, and $R$ is the transition system $TS= (S^P,\{s0\}^P,R{\uparrow}P)$.
\end{definition}
In other words, one can fully specify a multiagent transition system over $S$ and $P$ simply by providing a set of transactions over $S$, each with participants $Q\subseteq P$.

Given an arbitrary set of local states $\mathcal{S}$ with designated initial state $s0\in \mathcal{S}$, a \emph{local-states function} $S: P \mapsto 2^\mathcal{S}$ maps every set of agents $P \subset \Pi$ to some $S(P) \subset \mathcal{S}$ that includes $s0$ and satisfies $P \subset P' \subset \Pi \implies S(P) \subset S(P')$.  The symbol $S$ is overloaded: a set of local states in Definition~\ref{definition:at}, and the local-states function here, with $S(P)$ the set of local states for population $P$.

\begin{definition}[Multiagent Protocol]\label{definition:protocol}
A multiagent protocol $\mathcal{F}$ over a local-states function $S$ is a family of multiagent transition systems that has exactly one transition system 
$$\mathcal{F}(P) = (C(P),c0(P),T(P))$$
for every $P \subset \Pi$, where $C(P) := S(P)^P$, $c0(P) := \{s0\}^P$, and $T(P)$ is its transition relation.
\end{definition}

\begin{definition}[Multiagent Protocol Induced by Transactions]\label{definition:protocol-induced}
Let $S$ be a local-states function and $R$ a set of transactions, each over some $Q \subset \Pi$ and $S(Q)$. 
A multiagent protocol $\mathcal{F}$ is \emph{induced by $R$} if for each set of agents $P \subset \Pi$:
$$\mathcal{F}(P) = (S(P)^P, \{s0\}^P, R(P){\uparrow}P)$$
where $R(P) := \{ t \in R : t \text{ is over } Q \text{ and } S(Q) \text{ for some } Q \subseteq P\}$.
\end{definition}
Hence an agent whose state a transaction reads or names is one of its participants, active or stationary.

The same transaction---such as ``sync $p$ and $q$''---can be carried out in many configurations, yielding different transactions that are nonetheless ``the same action.''  An equivalence relation on transactions captures this.

\begin{definition}[Transaction Equivalence]\label{definition:equivalence}
Given a set of transactions $R$, a \temph{transaction equivalence} is an equivalence relation $\sim$ on $R$ such that $t \sim t'$ implies $t$ and $t'$ have the same participants.  We write $[t]$ for the equivalence class of $t$ under $\sim$.
\end{definition}
An equivalence class $[t]$ is \temph{enabled} in configuration $c$ if some $t' \in [t]$ is enabled in $c$.

For example, in the specifications of Section~\ref{section:specifications}, all Sync transactions over a given pair of agents---differing only in the feeds they synchronise---are equivalent: they represent ``the same action'' in different configurations.

\begin{definition}[Live and Correct Run]\label{definition:liveness}
Given a set of transactions $R$ with equivalence $\sim$ and a designated set $\Lambda \subseteq R/{\sim}$ of \temph{live} equivalence classes, a run $r$ is \temph{live} if for every $[t] \in \Lambda$: no suffix of $r$ has $[t]$ enabled throughout with no member of $[t]$ taken in the suffix.  A run is \temph{correct} if it is live.
\end{definition}
This is a fairness condition: a live equivalence class that remains enabled is eventually taken.  The set $\Lambda$ distinguishes obligatory transactions---which must eventually be carried out when persistently enabled---from voluntary ones, such as posting a message, which carry no liveness obligation.  Without this distinction, liveness would force every persistently-enabled Post to be executed.

\subsection{Essential Agents and Global Platform Classes}\label{subsection:characterisation}

The essential agents of a protocol capture its structural dependencies: which agents must be present for multi-party interaction to be possible?

\begin{definition}[Essential Agents]
Given a multiagent protocol $\mathcal{F}$ induced by transactions $R$, a set of agents $E \subseteq \Pi$ is \emph{essential} if it is a minimal set such that for every $P$ with $P\cap E = \emptyset$, every run of $\mathcal{F}(P)$ has only unary transitions. 
\end{definition}
The set $\Pi$ always satisfies the condition, since no nonempty $P \subset \Pi$ has $P \cap \Pi = \emptyset$; and as every $P$ is finite, the intersection of a chain of sets satisfying the condition satisfies it, so by Zorn's lemma an essential set exists and the minimum cardinality over all essential sets is well-defined.  Note that a protocol may have multiple essential sets of minimum cardinality.  A protocol has \emph{all agents but one essential} if every essential set $E$ with $\emptyset \subsetneq E \subsetneq \Pi$ has $|\Pi \setminus E| = 1$.

\begin{definition}[Global Platform Classes]\label{definition:classes}
Let $\calF$ be a transactions-based multiagent protocol admitting an essential set $E$ with $\emptyset \subsetneq E \subsetneq \Pi$, and let $\kappa$ be the minimum cardinality of such a set.  Then the \emph{class} of $\calF$ is defined as follows:
\begin{enumerate}
\item \textbf{Centralised}:  $\kappa=1$
\item \textbf{Decentralised}:  $1< \kappa< \infty$ 
\item \textbf{Federated}:  $\kappa = \infty$ and some essential set $E$ has $|\Pi \setminus E| \ge 2$
\item \textbf{Grassroots}: every essential set $E$ with $\emptyset \subsetneq E \subsetneq \Pi$ has $|\Pi \setminus E| = 1$
\end{enumerate}
\end{definition}

\begin{observation}
 The four classes of Definition~\ref{definition:classes} partition the transactions-based multiagent protocols admitting an essential set $E$ with $\emptyset \subsetneq E \subsetneq \Pi$.
\end{observation}

\begin{proof}
Let $\kappa$ be the minimum cardinality of an essential set $E$ with $\emptyset \subsetneq E \subsetneq \Pi$, which is well-defined.  If $\kappa < \infty$, a minimum essential set has $|\Pi \setminus E| = \infty$, so the Grassroots condition fails, and exactly one of Centralised ($\kappa = 1$) and Decentralised ($1 < \kappa < \infty$) holds.  If $\kappa = \infty$, neither of those holds, and exactly one of Federated (some essential set has $|\Pi \setminus E| \ge 2$) and Grassroots (every essential set has $|\Pi \setminus E| = 1$) holds, as these are complementary.  Hence exactly one class applies.
\end{proof}

Namely, to classify any global platform---existing or imagined---it is enough to provide a credible multiagent atomic transactions specification of it and analyse the cardinality of the essential set of the ensuing multiagent protocol.\footnote{It is theoretically possible for two alternative specifications to vie for being the ``right'' specification of some known global platform.  If the two seem credible yet have different cardinalities of the minimal sets of essential agents,  probably the specification with the larger set should be used for classification.  We have yet to see or produce such a case.}

The class of a platform depends on which agents its specification designates as servers, and that designation reflects a capability restriction: operating a fixed-address, always-on server requires a domain name and a permanently reachable host, which not every person has.  The essential sets are infinite in both the federated and the grassroots class, so this restriction separates the two classes and the number of servers does not.  Anyone may create a Mastodon instance, and every client depends on the instance it joined, so Mastodon is federated; were operating a server available to every agent, any two agents could serve each other and all agents but one would be essential.

In the grassroots class, $E = \Pi \setminus \{p\}$ for some $p$ means that every agent except one is essential.  The single excluded agent can only talk to oneself---any single agent alone can only perform unary transitions---so the essential set is as large as it can possibly be.  Conversely, no two agents can be excluded: in a grassroots protocol, any two agents can interact.  This essential-agents notion is the \emph{loose} characterisation; Section~\ref{section:grassroots} gives two stronger (\emph{tight}) definitions from the literature, each of which implies it.

\section{Global Social Networks: Specifications and Classification}\label{section:specifications}

Here we use the framework of Section~\ref{section:framework} to provide specifications for a Facebook/X-like social network with feeds and followers for each platform class, exposing their architectural differences. For each specification we state its fairness requirement (the designation of $\Lambda$), correctness, and the cardinality of essential agents; proofs appear in Appendix~\ref{appendix:proofs}.  In every specification below, Sync transactions are grouped by their participants into equivalence classes; all other transactions form singleton equivalence classes.  Throughout, posts are sequences over a fixed message alphabet $M$ and $\varepsilon$ is the empty sequence; lowercase $posts$ and ``$p$ follows $q$'' name projections of the local state, whereas capitalised Post and Follow name transactions.

\subsection{Correctness of a Social Network}
The social network we specify enables agents to post messages and follow other agents to receive their posts. Regardless of architecture, any specification of such network should guarantee two fundamental properties:

\begin{definition}[Social Network Correctness]\label{definition:sn-correctness}
A \emph{social network} is a multiagent protocol $\mathcal{F}$ induced by transactions $R$, equipped, for each $P \subset \Pi$, with a relation ``$p$ follows $q$'' on configurations of $\mathcal{F}(P)$ and a projection $posts_q$ within $c_p$ for $p, q \in P$, such that whenever $p$ follows $q$ in $c$ both $posts_q$ within $c_p$ and $posts_q$ within $c_q$ are defined. It satisfies \emph{Social Network Correctness} if:
\begin{enumerate}
    \item  \textbf{Follower Correctness}: For every $P \subset \Pi$, correct run $r$ of $\mathcal{F}(P)$, and all $p, q \in P$, if $p$ follows $q$ in configuration $c \in r$, then:
\begin{itemize}
\item \emph{Safety}: $posts_q$ within $c_p$ is a prefix of $posts_q$ within $c_q$.
\item \emph{Liveness}: For any post $m$ in $posts_q$ within $c_q$, eventually $m$ appears in $posts_q$ within $c_p$.
\end{itemize}
    \item \textbf{Agent Autonomy}: For all $P \subset \Pi$ and all runs of $\mathcal{F}(P)$:
\begin{itemize}
\item \emph{Post autonomy}: For any initialised agent $p \in P$ and any message $m$, Post$(m)$ is enabled.
\item \emph{Follow autonomy}: For any initialised agent $p \in P$ and $q \in P$ with $q \neq p$, if $p$ does not follow $q$, then a Follow transaction is enabled.
\end{itemize}
\end{enumerate}
\end{definition}

\subsection{Centralised Platforms}\label{subsection:centralised}

Centralised platforms employ a named server agent that all other agents must interact with. We illustrate this with a social network where a central server maintains feeds for all registered users. Users can post locally but need the server to exchange content.  For a post by $p$ to reach $q$ the following transactions have to take place: (1) $p$ registers with the server (2) $q$ registers with the server  (3) $q$ follows $p$ (4) $p$ posts (5) $p$ syncs with the server (6) $q$ syncs with the server.  Subsequent posting by $p$ require only the last three transactions.

\begin{definition}[Centralised Social Network Transactions]\label{definition:social-network}
Given agents $P \subset \Pi$ with a named server $p \in P$ and users $P \setminus \{p\}$. Local states are sets of feeds, where a feed is a pair $(agent, posts)$ with $posts$ a sequence. Initial local state is $\emptyset$.  The transactions are $c \rightarrow c'$ where:
\begin{itemize}
\item \textbf{Register} over $\{p,q\}$: 
  \begin{itemize}
  \item $c_q = \emptyset$ and $c'_q := \{(q, \varepsilon)\}$
  \item $c'_p := c_p \cup \{(q, \varepsilon)\}$
  \end{itemize}

\item \textbf{Post}$(m)$ over $\{q\}$:  $(q,posts) \in c_q$ and $c'_q := c_q \setminus \{(q,posts)\} \cup \{(q,posts\cdot m)\}$.

\item \textbf{Follow}$(q')$ over $\{q, q'\}$:
$c_q \ne \emptyset$, $q'$ has no feed in $c_q$, and $c'_q := c_q \cup \{(q', \varepsilon)\}$

\item \textbf{Sync} over $\{q,p\}$: $q$ has a feed in $c_p$ and:
  \begin{itemize}
  \item $c'_p := c_p$ with $q$'s posts missing in $c_p$ added to $(q, posts)$
  \item $c'_q := c_q$ with posts by followed agents missing in $c_q$ added
  \end{itemize}
\end{itemize}
\end{definition}

The transaction equivalence groups all Sync transactions over the same pair $\{q, p\}$ into one class.  The live set $\Lambda$ consists of these Sync classes: for any registered agent $q$ (i.e., $q$ has a feed in $c_p$), Sync over $\{q,p\}$ must eventually be taken.  Post and Follow are voluntary and not in $\Lambda$.

The server $p$ is required for any transaction to occur: Register and Sync have $p$ as a participant, and Post and Follow require registration.

Posts reach a follower through the server: the followed agent syncs to the server and the follower syncs from it, preserving order.

\begin{restatable}{proposition}{thmCentralisedCorrect}\label{theorem:centralised-correct}
The centralised social network protocol satisfies Social Network Correctness.
\end{restatable}

\begin{restatable}{proposition}{thmCentralisedEssential}\label{theorem:centralised-essential}
In the centralised social network protocol, $E = \{p\}$ and $|E| = 1$.
\end{restatable}

\subsection{Decentralised Platforms}\label{subsection:decentralised}

\mypara{Bitcoin} We illustrate decentralised platforms with bootstrap agents using an abstract Bitcoin-like protocol. In particular, we do not specify preconditions for the nondeterministic AddBlock  transaction.
While decentralised systems typically distinguish between full nodes and lightweight nodes, we abstract away this distinction since any node can in principle become a full node. The genesis block and bootstrap node addresses are public information hardcoded in the protocol.

\begin{definition}[Bitcoin Transactions]\label{definition:bitcoin}
Given a constant set of bootstrap agents $B \subset \Pi$ and agents $P \subset \Pi$. Local states $S$ are pairs (blockchain, peers) where a \emph{block} is opaque data (carrying posts in Definition~\ref{definition:decentralised-social}), blockchain is a sequence of blocks, and peers is a set of agents,
and for configuration $c$ over $S$ and $P$, we let $c_p = (chain_p, peers_p)$.
Initial local state is $(g, B)$ where $g$ is the genesis block.  Transitions include $c\rightarrow c'$ where:

\begin{itemize}
\item \textbf{AddBlock}$(b)$ over $\{p\}$:  $c'_p := (chain_p \cdot b, peers_p)$.

\item \textbf{Sync} over $\{p, q\}$:  $q \in peers_p$, $|chain_p| < |chain_q|$ and:
  \begin{itemize}
  \item $chain'_p := chain'_q := chain_q$, 
  \item $peers'_p := peers'_q := peers_p \cup peers_q \cup \{p\}$
  \end{itemize}
\end{itemize}
\end{definition}

If the chain of $p$ is not a prefix of the chain of $q$ upon a Sync over $\{p,q\}$, then 
the blocks in $p$ that are inconsistent with the chain in $q$ are \emph{abandoned} and we say that the chain of $p$ has undergone \emph{reorg}. The correctness of the protocol depends on the probability of an AddBlock transaction being much lower than the probability of a Sync transaction~\cite{garay2015bitcoin,pass2017analysis}. This in turn implies that the probability of a reorg abandoning a block diminishes quickly as a function of how deep the block is ``buried'', namely how many blocks follow it in the blockchain~\cite{nakamoto2008bitcoin,guo2022bitcoin}. Moreover, the probabilistic argument underscores the importance of the bootstrap nodes: It depends on them forming a connected component and on all active agents joining this component; the argument falls apart if multiple connected components form and grow independently~\cite{heilman2015eclipse,apostolaki2017hijacking}.

\mypara{Decentralised Social Network} Each agent maintains their own posts and publishes them directly to the blockchain. Chain reorganizations require rolling back the publication pointer for posts in abandoned blocks.  

For a post by $p$ to reach $q$ the following unbounded sequence of transactions have to take place: (1) $p$ posts block $b$ (2) a sequence of Sync transactions in which $b$ is part of the longer chain, the last of which is with $q$.
If a Sync transaction with $p$ entails a reorg in which the block $b$ is abandoned , it has to be posted again.

\begin{definition}[Decentralised Social Network Transactions]\label{definition:decentralised-social}
Given agents $P \subset \Pi$ with bootstrap nodes $B \subset P$. Local state is $(blockchain, peers, posts, pointer)$ where blockchain is a sequence of blocks, peers is a set of agents, posts is a sequence of posts, and pointer indexes published posts, and for configuration $c$ over $S$ and $P$, we let $c_p = (chain_p, peers_p,posts_p, pointer_p)$.  Initial local state is
$(g, B, \varepsilon, 0)$.  The transactions are  $c \rightarrow c'$ where:
\begin{itemize}
\item \textbf{Post}$(m)$ over $\{p\}$:  $c'_p := (chain_p, peers_p, posts_p \cdot m, pointer_p)$.

\item \textbf{AddBlock}$(b)$ over $\{p\}$:   $b$ contains posts from $pointer_p$ to $|posts_p|$, then $c'_p := (chain_p \cdot b, peers_p, posts_p, |posts_p|)$.

\item \textbf{Sync} over $\{p, q\}$: $q \in peers_p$, $|chain_p| < |chain_q|$ and:
  \begin{itemize}
   \item $chain'_p := chain'_q := chain_q$ 
  \item $peers'_p := peers'_q := peers_p \cup peers_q \cup \{p\}$
  \item  $pointer'_q:=pointer_q$, and $pointer'_p$ is the number of posts by $p$ contained in the blocks of $x$, where $x$ is the longest common prefix of $chain_p$ and $chain_q$
  \end{itemize}
\end{itemize}
\end{definition}

While several proposals exist for blockchain-based social networks~\cite{DSNP2021,buchegger2009peerson,chockler2007constructing,chockler2007spidercast}, our specification exposes fundamental architectural mismatches:
\begin{itemize}
\item \textbf{Replication}: All posts must be replicated to all agents and stored by all agents, regardless of who follows whom
\item \textbf{Consensus}: Every post requires global consensus.
\item \textbf{Instability}: Posts may be reordered after publication and responses may appear before their referents.
\end{itemize}
Blockchain efficiency can be improved by sharding~\cite{kokoris2018omniledger,wang2019monoxide}, layering~\cite{poon2016bitcoin,decker2015fast}, and side channels~\cite{dziembowski2018general,miller2017sprites}, but the issues raised above cannot be addressed short of operating the social network via a different architecture.  Perhaps the key reason for that is that social networks do not need consensus to operate (see the grassroots social network protocol below, Definition~\ref{definition:grassroots-social}), so there is no reason to pay the price for consensus in order to realise them.

The live set $\Lambda$ consists of the Sync classes: for any agents $p, q$ where $q \in peers_p$, Sync over $\{p,q\}$ must eventually be taken.  Post and AddBlock are not in $\Lambda$.

The bootstrap nodes $B$ are hardcoded in the initial state as the peer set of every agent; without them, no Sync is possible.

Every agent replicates the whole blockchain, so all posts reach everyone and follower correctness holds trivially.

\begin{restatable}{proposition}{thmDecentralisedCorrect}\label{theorem:decentralised-correct}
The decentralised social network protocol satisfies Social Network Correctness.
\end{restatable}

\begin{restatable}{proposition}{thmDecentralisedEssential}\label{theorem:decentralised-essential}
In the decentralised social network protocol, $E = B$ and $1 < |E| < \infty$.
\end{restatable}

\subsection{Federated Platforms}\label{subsection:federated}

\mypara{Federated Social Network} 
Multiple servers cooperate through federation. Users bind to home servers with qualified identities (e.g., p@s). Federation occurs when users follow remote users.

For a post by $p$ to reach $q$ the following eight transactions have to take place: (1) $p$ registers with server $r$ (2) $q$ registers with server $s$  (3) $q@s$ follows $p@r$  (4) $q@s$ syncs with the server $s$. (5) $p@r$ posts (6) $p@r$ syncs with the server $r$ (7) server $r$ syncs with server $s$ (8) $q@s$ syncs with server $s$.
Subsequent posts by $p$ require only the last four transactions.

\begin{definition}[Federated Social Network Transactions]\label{definition:federated}
Given agents $P \subset \Pi$ with designated servers $Q \subset P$ and clients $P \setminus Q$. Local states are sets of feeds, where a feed is $(agent@server, posts)$ with $posts$ a sequence. 
Initial local state is $\emptyset$. Transactions are $c\rightarrow c'$ where:
\begin{itemize}
\item \textbf{Register} over $\{p,s\}$: $c_p = \emptyset$, $c'_p := \{(p@s, \varepsilon)\}$, and $c'_s := c_s \cup \{(p@s, \varepsilon)\}$

\item \textbf{Post}$(m)$ over $\{p\}$: $(p@s,posts) \in c_p$ and $c'_p := c_p \setminus \{(p@s,posts)\} \cup \{(p@s,posts\cdot m)\}$.

\item \textbf{Follow}$(q@s')$ over $\{p, q, s'\}$:
$c_p \ne \emptyset$, $q@s'$ has no feed in $c_p$, and $c'_p := c_p \cup \{(q@s', \varepsilon)\}$

\item \textbf{Sync} over $\{a,b\}$: 
  \begin{itemize}
  \item If $a \in P \setminus Q$ and $b \in Q$ where $a@b$ has a feed in $c_b$: 
    \begin{itemize}
    \item $c'_b := c_b$ with $(a@b, posts)$ updated to include any posts from $c_a$
    \item $c'_a := c_a$ with feeds of followed users updated from $c_b$'s copies
    \end{itemize}
  \item If $a, b \in Q$ and $a \ne b$: For each $p@a$ with a feed in $c_b$, or $q@b$ with a feed in $c_a$:
    \begin{itemize}
    \item If $(p@a, posts) \in c_a$ and $(p@a, posts') \in c_b$: update to longer sequence in both
    \item If $(q@b, posts) \in c_b$ and $(q@b, posts') \in c_a$: update to longer sequence in both
    \end{itemize}
  \end{itemize}
\end{itemize}
\end{definition}

The live set $\Lambda$ consists of two kinds of Sync classes: (\ia) client-server: for any registered client $p$ at server $s$ (i.e., $p@s$ has a feed in $c_s$), Sync over $\{p,s\}$ must eventually be taken; (\ib) server-server: for any servers $s_1, s_2$ where $p@s_1$ has a feed in $c_{s_2}$ for some agent $p$, Sync over $\{s_1,s_2\}$ must eventually be taken.

Servers are required for any transaction to occur: Register requires a server, client-server and server-server Sync both require at least one server as participant, and Post and Follow require registration.  Without any server, no client can initialise.

The designation of $Q$ is the capability restriction of Section~\ref{subsection:characterisation}: operating a fixed-address, always-on server is not available to every agent.  Were it available to every agent, the same transactions with $Q = P$ would specify a grassroots platform, with all agents but one essential.

Posts propagate from a client to its server, between servers, and to the following client's server, preserving order along the way.

\begin{restatable}{proposition}{thmFederatedCorrect}\label{theorem:federated-correct}
The federated social network protocol satisfies Social Network Correctness.
\end{restatable}

\begin{restatable}{proposition}{thmFederatedEssential}\label{theorem:federated-essential}
In the federated social network protocol, $E = Q$, $|E| = \infty$, and $|\Pi \setminus E| \ge 2$.
\end{restatable}

\subsection{Grassroots Platforms}\label{subsection:grassroots}

\mypara{Grassroots Social Network} 
For a post by $p$ to reach $q$ the following five transactions have to take place: (1) $p$ initialises (2) $q$ initialises  (3) $q$ follows $p$ (4) $p$ posts (5) $p$ syncs with $q$.   Subsequent posting by $p$ require only the last  two transactions.

\begin{definition}[Grassroots Social Network Transactions]\label{definition:grassroots-social}
Given agents $P \subset \Pi$. Local states are sets of feeds $(agent, posts)$, where $posts$ is a sequence. Initial local state is $\emptyset$. Transactions are  $c \rightarrow c'$ where:
\begin{itemize}
\item \textbf{Initialise} over $\{p\}$: $c_p = \emptyset$ and $c'_p := \{(p, \varepsilon)\}$

\item \textbf{Post}$(m)$ over $\{p\}$:  $(p, s) \in c_p$ and $c'_p := c_p \setminus \{(p, s)\} \cup \{(p, s \cdot m)\}$

\item \textbf{Follow}$(q)$ over $\{p, q\}$:  $p\ne q$, $q$ has no feed in $c_p$, $c'_p := c_p \cup \{(q, \varepsilon)\}$  

\item \textbf{Sync} over $\{p, q\}$: for any $r$ with $(r, s) \in c_p$ and $(r, s') \in c_q$:
  \begin{itemize}
  \item If $|s| < |s'|$: update $(r, s)$ to $(r, s')$ in $c'_p$
  \item If $|s'| < |s|$: update $(r, s')$ to $(r, s)$ in $c'_q$
  \end{itemize}
\end{itemize}
\end{definition}

The live set $\Lambda$ consists of the Sync classes: for any agents $p, q$ where $q$ has a feed in $c_p$, Sync over $\{p,q\}$ must eventually be taken.  Initialise, Post, and Follow are voluntary and not in $\Lambda$.

A follower obtains posts by direct Sync with the followed agent, preserving order, with no intermediary.

\begin{restatable}{proposition}{thmGrassrootsCorrect}\label{theorem:grassroots-correct}
The grassroots social network protocol satisfies Social Network Correctness.
\end{restatable}

\begin{restatable}{proposition}{thmGrassrootsEssential}\label{theorem:grassroots-essential}
In the grassroots social network protocol, $E = \Pi \setminus \{p\}$ for some $p \in \Pi$.
\end{restatable}

\section{Grassroots Platforms: Two Characterisations}\label{section:grassroots}

Grassroots platforms have two previous definitions~\cite{shapiro2023grassrootsBA,shapiro2025atomic,lewis2026volitional}.  Here we present both, prove them incomparable, prove that each implies all agents but one are essential, placing grassroots platforms in the fourth class of Definition~\ref{definition:classes}, and prove that both implications are strict.

\subsection{The Original Definition}\label{subsection:grassroots-original}

The original definition of grassroots protocols~\cite{shapiro2023grassrootsBA,shapiro2025atomic} is based on the notion of interactivity.

\begin{definition}[Projection]\label{definition:projection}
For a configuration $c \in S^{P'}$ and subset $P \subseteq P'$, the \emph{projection} of $c$ to $P$, denoted $c/P$, is the configuration in $S^P$ defined by $(c/P)_p = c_p$ for all $p \in P$.
\end{definition}

\begin{definition}[Computation Notation]\label{definition:computation-notation}
For configurations $c, c'$ in a transition system, we write $c \xrightarrow{*} c'$ if there exists a finite computation (sequence of transitions) from $c$ to $c'$. Specifically, there exist configurations $c = c_0, c_1, \ldots, c_n = c'$ where $c_i \rightarrow c_{i+1}$ is a transition for each $i < n$.
\end{definition}

\begin{definition}[Interactive Protocol]\label{definition:interactive}
A protocol $\mathcal{F}$ is \emph{interactive} if for every $\emptyset \subset P \subset P' \subseteq \Pi$ and every configuration $c \in C(P')$ such that $c/P \in C(P)$, there is a computation $c \xrightarrow{*} c'$ of $\mathcal{F}(P')$ for which $c'/P \notin C(P)$.
\end{definition}

The interactive property requires that agents in $P$ can always potentially interact with agents in $P' \setminus P$, leaving ``alien traces'' in their local states that could not have been produced by $P$ operating alone.

\begin{definition}[Grassroots Protocol (Original)]\label{definition:grassroots-original}
An atomic transactions-based protocol $\mathcal{F}$ is \emph{grassroots} if it is interactive.
\end{definition}
We note that the original definition of grassroots protocols~\cite{shapiro2023grassrootsBA} included a notion of obliviousness, however subsequent work~\cite{shapiro2025atomic} showed that any transactions-based multiagent protocol is oblivious. Hence we bypass it here.

\begin{theorem}\label{theorem:interactive-essential}
In a grassroots protocol (under the original definition) all agents but one are essential.
\end{theorem}
\begin{proof}
Assume $\mathcal{F}$ is grassroots, then by definition it is interactive. Suppose for contradiction that some essential set $E$ with $|\Pi \setminus E| \ge 2$, so there exist at least two agents $p \ne q$ with $p, q \notin E$.

Let $P = \{p\}$ and $P' = \{p,q\}$. Note that $P \subset P'$ and $P' \cap E = \emptyset$.

Since $E$ is essential and $P' \cap E = \emptyset$, every run of $\mathcal{F}(P')$ has only unary transitions.  A unary transition of $\calF(P')$ active at $p$ is a transition of $\calF(P)$, as its transaction is over $\{p\}$ and $S(\{p\})$ and so is in $R(P)$; a transition active at $q$ leaves $p$'s state unchanged.  Hence for every run $c0(P') \xrightarrow{*} c'$ of $\calF(P')$, the trace of $p$'s state is reproducible by a run of $\calF(P)$, so $c'/P \in C(P)$.

But $\mathcal{F}$ is interactive, so for $P = \{p\} \subset P' = \{p,q\}$ and $c = c0(P')$, which has $c/P = c0(P) \in C(P)$, there is a computation $c \xrightarrow{*} c'$ of $\mathcal{F}(P')$ with $c'/P \notin C(P)$; it is a run of $\mathcal{F}(P')$, contradicting the previous paragraph.
\end{proof}

\subsection{The Interleaving-Based Definition}\label{subsection:grassroots-interleaving}

A recent alternative definition~\cite{lewis2026volitional} captures the informal notion of grassroots directly: two disjoint groups of agents can each operate independently---their interleaved correct runs are correct runs of the combined system---yet the combined system offers genuinely new behaviours that neither group could produce on its own.  To distinguish the three interleaving-based notions from their original counterparts, we mark them with a prime: $\text{oblivious}'$, $\text{interactive}'$, $\text{grassroots}'$.

\begin{definition}[Interleaving]\label{definition:interleaving}
Let $P, P' \subset \Pi$ be disjoint nonempty sets of agents, $r = c_0, c_1, \ldots$ a run of $\calF(P)$, and $r' = d_0, d_1, \ldots$ a run of $\calF(P')$.  An \temph{interleaving} of $r$ and $r'$ is a sequence $e_0, e_1, \ldots$ of configurations in $C(P \cup P')$ for which there exist non-decreasing sequences of indices $(i_k)_{k \geq 0}$ and $(j_k)_{k \geq 0}$ with $i_0 = j_0 = 0$ such that for every $k \geq 0$:
\begin{enumerate}
\item $(e_k)_p = (c_{i_k})_p$ for every $p \in P$,
\item $(e_k)_q = (d_{j_k})_q$ for every $q \in P'$,
\item if $e_{k+1}$ exists, then exactly one of:
  (\ia) $i_{k+1} = i_k + 1$ and $j_{k+1} = j_k$ (a $P$-step), or
  (\ib) $i_{k+1} = i_k$ and $j_{k+1} = j_k + 1$ (a $P'$-step).
\end{enumerate}
Moreover, if $r$ is finite of length $n$ then $i_k = n$ for some $k$, and if $r$ is infinite then for every $m \ge 0$ there is a $k$ with $i_k = m$; likewise for $r'$ and $(j_k)$.
\end{definition}

\begin{definition}[Interaction]\label{definition:interaction}
Let $Q \subset \Pi$.  A transition $e \to e'$ of $\calF(Q)$ is an \temph{interaction between} $x \ne y \in Q$ if $e/\{x\} \ne e'/\{x\}$ and $e/\{y\} \ne e'/\{y\}$, and $e/\{x\} \to e'/\{x\}$ is not a transition of $\calF(\{x\})$ or $e/\{y\} \to e'/\{y\}$ is not a transition of $\calF(\{y\})$.  It is an \temph{interaction between} disjoint nonempty $P, P' \subseteq Q$ if it is an interaction between some $x \in P$ and some $y \in P'$.
\end{definition}

\begin{definition}[Oblivious$'$, Interactive$'$, Grassroots$'$]\label{definition:grassroots-interleaving}
A  protocol $\calF$ is:
\begin{enumerate}
    \item \temph{oblivious$'$} if for every disjoint nonempty $P, P' \subset \Pi$,
    every interleaving of a correct run of $\calF(P)$ and a correct run of $\calF(P')$ is a correct run of $\calF(P\cup P')$.
    \item  \temph{interactive$'$} if for every $Q \subset \Pi$ and disjoint nonempty $P, P' \subseteq Q$, every prefix of a correct run of $\calF(Q)$ containing no interaction between $P$ and $P'$ is a prefix of a correct run of $\calF(Q)$ that contains one.
    \item \temph{grassroots$'$} if it is oblivious$'$ and interactive$'$.
\end{enumerate}
\end{definition}

Being oblivious$'$ means that two disjoint groups of agents, each running the protocol independently and correctly, do not interfere with each other.  Being interactive$'$ means that two groups that have not yet interacted can always still do so: whatever they have done apart, the combined system has a correct continuation in which a single transition engages agents of both, which no interleaving of their separate runs can produce.

\begin{theorem}\label{theorem:interleaving-essential}
In an interactive$'$ protocol all agents but one are essential.
\end{theorem}
\begin{proof}
Suppose for contradiction that some essential set $E$ has $|\Pi \setminus E| \ge 2$.  Then there exist at least two agents $p \ne q$ with $p, q \notin E$.  Let $Q = \{p,q\}$, $P = \{p\}$ and $P' = \{q\}$.  Since $Q \cap E = \emptyset$ and $E$ is essential, every run of $\calF(Q)$ has only unary transitions, so no transition of such a run changes the local states of both $p$ and $q$, and by Definition~\ref{definition:interaction} none is an interaction between $P$ and $P'$.  Hence the empty prefix of a correct run of $\calF(Q)$ contains no interaction between $P$ and $P'$ and is a prefix of no correct run that contains one, contradicting the interactivity$'$ of $\calF$.
\end{proof}

Since grassroots$'$ protocols are interactive$'$, the same holds for them.

\mypara{Relation between the two definitions}
The two definitions of grassroots are incomparable in general---neither implies the other---but for all platforms studied thus far, both definitions agree.  The interleaving-based definition~\cite{lewis2026volitional} is simpler and better captures the informal notion of grassroots: multiple instances can form and operate independently, yet may coalesce when interconnected.

Two protocols separate the definitions, one in each direction.

\begin{definition}[Follow Only]\label{definition:followonly}
Given agents $P \subset \Pi$.  A local state is a set $F \subseteq P$ of the agents followed; the initial local state is $\emptyset$.  The one transaction is:
\begin{itemize}
\item \textbf{Follow}$(q)$ over $\{p, q\}$: $p \ne q$ and $q \notin F_p$; $F'_p := F_p \cup \{q\}$
\end{itemize}
The live set $\Lambda$ is empty.
\end{definition}

\begin{definition}[Paired Counter]\label{definition:counter}
Given agents $P \subset \Pi$.  A local state is a pair $(n,a)$ with $n \in \mathbb{N}$ and $a \in P \cup \{\bot\}$; the initial local state is $(0,\bot)$.  The one transaction is:
\begin{itemize}
\item \textbf{Tick} over $\{p,q\}$: $n'_p := n_p + 1$ and $n'_q := n_q + 1$; $a_p$ and $a_q$ unchanged
\end{itemize}
The live set $\Lambda$ is empty.  No transaction changes the second component of a local state; it is there so that the local-states function is strictly monotone.
\end{definition}

\begin{restatable}{proposition}{thmIncomparable}\label{theorem:incomparable}
Follow Only is interactive and not interactive$'$, and Paired Counter is interactive$'$ and not interactive.  Hence neither definition implies the other.
\end{restatable}

Both definitions are \emph{tight}: each implies the \emph{loose} characterisation of Definition~\ref{definition:classes}---all agents but one essential---but, as discussed below, not conversely.  The question arises, which platforms qualify as grassroots under the loose characterisation but under neither tight definition?  As far-fetched as it may sound, we believe these are platforms in which participants sacrifice their dignity and forgo their basic human rights.  In one platform participants forgo their freedom of speech, and runs have a suffix in which no two agents communicate.  In another, after a finite prefix a leader is chosen who cannot be deposed and who thereafter mediates all agent-to-agent communication.  We specify both and prove that each has all agents but one essential, but is grassroots under neither tight definition.  The mathematical and practical implications of the difference between the tight and loose grassroots notions remains to be studied.

\begin{definition}[Silenced Speech]\label{definition:silenced}
Given agents $P \subset \Pi$.  A local state is a pair $(v,M)$ with $v \in \{\mathit{open},\mathit{silent}\}$ and $M \subseteq P$ the agents spoken with; the initial local state is $(\mathit{open},\emptyset)$.  Transactions are:
\begin{itemize}
\item \textbf{Speak} over $\{p,q\}$: $v_p = v_q = \mathit{open}$ and $q \notin M_p$; $M'_p := M_p \cup \{q\}$ and $M'_q := M_q \cup \{p\}$
\item \textbf{Silence} over $\{p\}$: $v_p = \mathit{open}$; $v'_p := \mathit{silent}$
\end{itemize}
The live set $\Lambda$ is empty: both transactions are voluntary.
\end{definition}

\begin{definition}[Entrenched Leader]\label{definition:leader}
Given agents $P \subset \Pi$.  A local state is a pair $(\ell,M)$ with $\ell \in P \cup \{\bot\}$ the agent's leader and $M \subseteq P$ the agents communicated with; the initial local state is $(\bot,\emptyset)$.  Transactions are:
\begin{itemize}
\item \textbf{Meet} over $\{p,q\}$: $\ell_p = \ell_q = \bot$ and $q \notin M_p$; $M'_p := M_p \cup \{q\}$ and $M'_q := M_q \cup \{p\}$
\item \textbf{Elect}$(r)$ over $\{p, r\}$: $\ell_p = \bot$; $\ell'_p := r$
\item \textbf{Relay} over $\{p,q\}$: $\ell_p = q$ and $q \notin M_p$; $M'_p := M_p \cup \{q\}$ and $M'_q := M_q \cup \{p\}$
\end{itemize}
The live set $\Lambda$ is empty.
\end{definition}
Agents communicate freely until they elect, and an election cannot be undone; thereafter an agent communicates only with its leader.

\begin{restatable}{proposition}{thmGapWitnesses}\label{theorem:gap}
In each of Silenced Speech and Entrenched Leader all agents but one are essential, and neither is interactive nor interactive$'$.
\end{restatable}

We can now show that the grassroots social network protocol (Definition~\ref{definition:grassroots-social}) satisfies both definitions.

\begin{restatable}{proposition}{thmGrassrootsIsGrassroots}\label{theorem:grassroots-is-grassroots}
The grassroots social network protocol is grassroots under the original definition and grassroots$'$ under the interleaving-based definition.
\end{restatable}

\section{Related Work}\label{section:related}

The characterisation of distributed systems has been a fundamental challenge in computer science since the field's inception. Our work builds upon and extends several research threads: formal models of distributed computation, atomic transactions in distributed systems, peer-to-peer architectures, and the emerging theory of grassroots computing.

\subsection{Formal Models of Distributed Systems}

Lynch's comprehensive treatment of distributed algorithms~\cite{lynch1996distributed} provides foundational definitions for distributed systems, though primarily focused on consensus and synchronization rather than architectural classification. Tel's work~\cite{tel2000introduction} similarly emphasizes algorithmic aspects without addressing the structural characterisation we identify. Attiya and Welch~\cite{attiya2004distributed} present a computational model based on message passing and shared memory, but do not address the architectural classification of platforms.

The closest work to ours in spirit is Angluin's seminal paper on local and global properties in networks of processors~\cite{angluin1980local}, which distinguishes between problems solvable with purely local information versus those requiring global coordination. However, Angluin focuses on computational complexity rather than architectural characterisation, and does not address the spectrum from centralized to grassroots that we formalize.

The notion of essential agents is analogous to other threshold concepts in distributed computing, such as the FLP impossibility result requiring failure detectors for consensus, or Byzantine agreement requiring $n > 3f$ participants.

\subsection{Atomic Transactions and Distributed Computing}

Atomic transactions have been extensively studied in distributed database systems~\cite{bernstein1987concurrency,gray1993transaction,weikum2001transactional}. The integration of atomic transactions into distributed computing models was pioneered by Lynch and colleagues~\cite{lynch1988atomic}, though their focus was on correctness properties rather than using transactions as a basis for system classification.

More recent work has explored atomic transactions in the context of blockchain systems~\cite{herlihy2018atomic}, though this literature assumes a specific architectural model (blockchain) rather than using transactions to characterize different architectures. The extension of process calculi with atomic transactions~\cite{acciai2007concurrent,de2010communicating} provides formal semantics but has not been applied to classify system architectures.

\subsection{Peer-to-Peer and Decentralized Systems}

The peer-to-peer systems literature provides extensive practical examples but limited formal characterisation. Lua et al.~\cite{lua2005survey} survey P2P overlay networks, categorizing them by topology (structured vs. unstructured) rather than by essential agent requirements. Androutsellis-Theotokis and Spinellis~\cite{androutsellis2004survey} classify P2P systems by application domain without addressing the fundamental architectural distinctions we identify.

Scuttlebutt~\cite{tarr2019secure} appears to be grassroots, though it has not been formally proven to be so. BitTorrent~\cite{cohen2003incentives}, the Dat protocol~\cite{ogden2017dat} and IPFS~\cite{benet2014ipfs} occupy an intermediate position, requiring bootstrap nodes similar to our characterisation of decentralized platforms.

\section{Conclusion}\label{section:conclusion}

We introduced essential agents---minimal sets whose removal prevents all non-unary transitions---and proved their cardinality partitions global platforms into four classes: centralised (1), decentralised (finite >1), federated (infinite, not universal), and grassroots (universal). We proved that the two previous definitions of grassroots platforms are incomparable, that each implies all agents but one are essential, and that both implications are strict. This work provides the first mathematical framework for studying today's most important family of computer systems.

\mypara{Use of AI tools} An AI assistant was used throughout the preparation of this paper: for discussion of the definitions and the proofs, for drafting and revising prose, and for checking citations.  The authors are accountable for all of its content.


\bibliography{bib}
\appendix

\section{Proofs}\label{appendix:proofs}

\thmCentralisedCorrect*
\begin{proof}
\textbf{Follower Correctness.}
\emph{Safety}: If $(q, posts) \in c_p$, then either (\ia) $p$ has not synced with server $s$ since following $q$, in which case $posts = \varepsilon$, or (\ib) $p$ has synced, receiving $q$'s posts from $c_s$. Since the server only appends to $(q, posts)$ in $c_s$ during Sync with $q$, and $p$ receives these during Sync with $s$, $posts$ in $c_p$ is always a prefix of $posts$ in $c_q$.
\emph{Liveness}: If $p$ follows $q$ and $q$ has posted $m$, then by liveness of the Sync class for $\{q,s\}$, Sync over $\{q,s\}$ eventually occurs, adding $m$ to $(q, posts)$ in $c_s$. Subsequently, Sync over $\{p,s\}$ occurs, copying the updated posts to $c_p$.

\textbf{Agent Autonomy.}
Post$(m)$ over $\{q\}$ requires only $(q, posts) \in c_q$, which holds after registration.
Follow$(q')$ over $\{q\}$ requires $c_q \neq \emptyset$ and $q'$ has no feed in $c_q$, always enabled for unfollowed agents.
\end{proof}

\thmCentralisedEssential*
\begin{proof}
Without the server ($p \notin P$), Register and Sync are unavailable, as both have $p$ as a participant; so no agent registers, and Post and Follow, which require a registered agent, are never enabled.  Hence $\{p\}$ is essential.  Minimality: the empty set is not essential since Sync over $\{q,p\}$ is a non-unary transition enabled when $p \in P$.  Therefore $E = \{p\}$ and $|E| = 1$.
\end{proof}

\thmDecentralisedCorrect*
\begin{proof}
\textbf{Follower Correctness.}
This holds vacuously as there is no Follow transaction. All agents receive all posts via blockchain replication.

\textbf{Agent Autonomy.}
Post$(m)$ over $\{p\}$ is always enabled as it only appends to $posts_p$.
Follow autonomy is not applicable (no Follow transaction defined).
\end{proof}

\thmDecentralisedEssential*
\begin{proof}
The initial local state is $(g, B)$, so $peers_p = B$ for every agent $p$.  If $P \cap B = \emptyset$, then for every $p \in P$ and every $q \in peers_p = B$, $q \notin P$, so Sync over $\{p,q\}$ is never enabled.  Only AddBlock and Post (both unary) remain.  Hence $B$ is essential.

Minimality: for any $b \in B$, consider $P = \{q, b\}$ where $q \notin B$.  Since $b \in peers_q = B$, Sync over $\{q, b\}$ is enabled---a non-unary transition.  Hence $B \setminus \{b\}$ does not suffice, and $E = B$ is minimal.  No single agent is essential: for any $a$, taking a client $q$ and a bootstrap node $b \in B$ with $q,b \neq a$ (possible as $|B|>1$), Sync over $\{q,b\}$ is enabled while $a \notin \{q,b\}$.  Hence $\kappa > 1$, and since $B$ is essential and finite, $1 < \kappa < \infty$---the protocol is decentralised.
\end{proof}

\thmFederatedCorrect*
\begin{proof}
\textbf{Follower Correctness.}
\emph{Safety}: If $(q@s', posts) \in c_p$, then $posts$ was obtained via Sync with $p$'s home server, which obtained it via federation from $s'$ (if $s' \neq s$) or directly (if $s' = s$). Server-to-server sync preserves prefix ordering, so $posts$ is a prefix of $q$'s actual posts.
\emph{Liveness}: If $p$ follows $q@s'$ and $q$ posts $m$: (\ia) by liveness of the client-server Sync class, Sync over $\{q,s'\}$ occurs, updating $(q@s', posts)$ in $c_{s'}$; (\ib) if $p$'s home server $s \neq s'$, then by liveness of the server-server Sync class, Sync over $\{s,s'\}$ occurs, updating $(q@s', posts)$ in $c_s$; (\ic) Sync over $\{p,s\}$ occurs, delivering $m$ to $p$.

\textbf{Agent Autonomy.}
Post$(m)$ over $\{p\}$ requires $(p@s, posts) \in c_p$, established at registration.
Follow$(q@s')$ over $\{p\}$ requires $c_p \neq \emptyset$ and $q@s'$ has no feed in $c_p$, enabled for unfollowed agents.
\end{proof}

\thmFederatedEssential*
\begin{proof}
If $P \cap Q = \emptyset$ (only clients, no servers), then the initial local state is $\emptyset$ and Register over $\{p,s\}$ requires $s \in Q$, which is unavailable.  Hence no client can initialise ($c_p$ remains $\emptyset$), and Post requires $p@s$ to have a feed in $c_p$, which never holds.  Follow requires $c_p \neq \emptyset$, so it too is never enabled.  No non-unary transition is possible.  Hence $Q$ is essential.

Minimality: for any server $s \in Q$, consider $P = \{q, s\}$ where $q \notin Q$.  Register over $\{q, s\}$ is enabled---a non-unary transition.  Hence $Q \setminus \{s\}$ does not suffice, and $E = Q$ is minimal.  Every essential set must contain all of $Q$, since each non-unary transaction has a server as a participant; hence every essential set is infinite and $\kappa = \infty$.  As the essential set $Q$ has $|\Pi \setminus Q| \ge 2$, the protocol is federated.
\end{proof}

\thmGrassrootsCorrect*
\begin{proof}
\textbf{Follower Correctness.}
\emph{Safety}: If $(q, posts) \in c_p$, then $posts$ was obtained via direct Sync with $q$ or transitively through other agents. The Sync transaction only updates to longer sequences, preserving prefix ordering.
\emph{Liveness}: If $p$ follows $q$ ($q$ has a feed in $c_p$) and $q$ posts $m$, then by liveness of the Sync class for $\{p,q\}$, Sync over $\{p,q\}$ eventually occurs, updating $(q, posts)$ in $c_p$ to include $m$.

\textbf{Agent Autonomy.}
Post$(m)$ over $\{p\}$ requires $(p, s) \in c_p$, established by Initialise.
Follow$(q)$ over $\{p\}$ requires $p \neq q$ and $q$ has no feed in $c_p$, enabled for unfollowed agents.
\end{proof}

\thmGrassrootsIsGrassroots*
\begin{proof}
We first show the protocol is interactive (Definition~\ref{definition:grassroots-original}).  Let $\emptyset \subset P \subset P' \subseteq \Pi$ and let $c \in C(P')$ with $c/P \in C(P)$.  Pick any $q \in P' \setminus P$ and any $p \in P$.  From $c$, agent $p$ can execute Initialise (if not yet initialised), then Follow$(q)$, producing a configuration in which $(q, \varepsilon) \in c'_p$.  No transaction in $\calF(P)$ can produce a feed for an agent outside $P$, so $c'/P \notin C(P)$.  Hence $\calF$ is interactive, and therefore grassroots under the original definition.

The protocol is also grassroots$'$ under the interleaving-based definition (Definition~\ref{definition:grassroots-interleaving}).  Obliviousness$'$ holds because the only transaction changing two local states is Sync, which requires that $r$ have a feed in both $c_p$ and $c_q$---and in an interleaving of runs of disjoint groups $P$ and $P'$, no agent in $P$ can have a feed for an agent in $P'$ (since Follow only adds feeds for agents named in its argument, and Sync only propagates existing feeds).  For interactivity$'$, let $Q \subset \Pi$, let $P, P' \subseteq Q$ be disjoint and nonempty, let $\hat{r}$ be a prefix of a correct run of $\calF(Q)$ containing no interaction between $P$ and $P'$, and pick $p \in P$ and $q \in P'$.  Extend $\hat{r}$ by whichever of Initialise, Follow and Post $p$ and $q$ have not already taken, so that each follows the other and each has posted, and then by Sync over $\{p,q\}$.  That Sync updates $p$'s feed for $q$ and $q$'s feed for $p$, so it changes the local states of both; and $p$'s state then holds a feed for $q \ne p$, which is not a local state of $\calF(\{p\})$, so its restriction to $\{p\}$ is not a transition of $\calF(\{p\})$.  It is therefore an interaction between $p$ and $q$, hence between $P$ and $P'$.  Extending further by taking every enabled Sync class in turn gives a correct run of $\calF(Q)$ having $\hat{r}$ as a prefix.  By Theorem~\ref{theorem:interleaving-essential}, all agents but one are essential.
\end{proof}

\thmGrassrootsEssential*
\begin{proof}
By Proposition~\ref{theorem:grassroots-is-grassroots}, the protocol is grassroots under both definitions.  By Theorems~\ref{theorem:interactive-essential} and~\ref{theorem:interleaving-essential}, all agents but one are essential, so $E = \Pi \setminus \{p\}$ for some $p \in \Pi$.
\end{proof}

\thmIncomparable*
\begin{proof}
\textbf{Follow Only is interactive.}  Let $\emptyset \subset P \subset P' \subseteq \Pi$ and $c \in C(P')$ with $c/P \in C(P)$, and take $p \in P$ and $q \in P' \setminus P$.  As $c/P \in C(P)$ we have $F_p \subseteq P$, so $q \notin F_p$ and Follow$(q)$ over $\{p\}$ is enabled at $c$; it yields $c'$ with $q \in F'_p$, so $c'/P \notin C(P)$.

\textbf{Follow Only is not interactive$'$.}  Its one transaction has a single active participant, so every transition of every run changes the local state of exactly one agent and none is an interaction between any two disjoint sets.  The empty prefix of a correct run therefore contains no interaction and is a prefix of no run that contains one.

\textbf{Paired Counter is interactive$'$.}  Let $Q \subset \Pi$, let $P, P' \subseteq Q$ be disjoint and nonempty, let $\hat r$ be a prefix of a correct run of $\calF(Q)$ containing no interaction between $P$ and $P'$, and take $p \in P$ and $q \in P'$.  Tick over $\{p,q\}$ is enabled at every configuration, and extending $\hat r$ by it gives a correct run, as $\Lambda$ is empty.  It changes the local states of both $p$ and $q$, and $\calF(\{p\})$ has no transitions, as Tick has two participants; so its restriction to $\{p\}$ is not a transition of $\calF(\{p\})$ and the Tick is an interaction between $p$ and $q$, hence between $P$ and $P'$.

\textbf{Paired Counter is not interactive.}  No transaction changes the second component of a local state, and the first ranges over $\mathbb{N}$ for every population, so $c/P \in C(P)$ implies $c'/P \in C(P)$ for every computation $c \xrightarrow{*} c'$.  Taking any $\emptyset \subset P \subset P' \subseteq \Pi$ and any such $c$ therefore refutes interactivity.
\end{proof}

\thmGapWitnesses*
\begin{proof}
\textbf{All agents but one essential.}  In both protocols the binary transaction---Speak, Meet---is enabled at the initial configuration for any two agents, where $v = \mathit{open}$, $\ell = \bot$ and $M = \emptyset$.  So a set $E$ omitting two agents $p \ne q$ is not essential, as the run of $\calF(\{p,q\})$ taking that transaction has a binary transition.  Hence every essential set omits at most one agent.  And $E = \Pi \setminus \{p\}$ is essential: $P \cap E = \emptyset$ gives $P \subseteq \{p\}$, so every transition of $\calF(P)$ is unary; and it is minimal by the preceding sentence.

\textbf{Not interactive.}  For Silenced Speech, let $\emptyset \subset P \subset P' \subseteq \Pi$ and let $c \in C(P')$ have $v_a = \mathit{silent}$ and $M_a = \emptyset$ for every $a \in P'$.  Then $c/P \in C(P)$, and no transaction is enabled at $c$, as Speak requires two open agents and Silence one.  So $c \xrightarrow{*} c'$ implies $c' = c$ and $c'/P \in C(P)$.

For Entrenched Leader, let $\emptyset \subset P \subset P' \subseteq \Pi$, fix $\ell \in P$, and let $c \in C(P')$ have $\ell_p = \ell$ and $M_p = P$ for every $p \in P$, and $\ell_a = a$ and $M_a = \emptyset$ for every $a \in P' \setminus P$.  Then $c/P \in C(P)$.  Meet and Elect require $\bot$, which no agent has; Relay over $\{a,b\}$ requires $\ell_a = b$ with $b \ne a$, which holds only for $a \in P$ and $b = \ell$, where $\ell \in M_a$ already.  So nothing is enabled at $c$ and $c'/P \in C(P)$ for every computation from $c$.  Neither protocol is therefore interactive.

\textbf{Not interactive$'$.}  For Silenced Speech, take any $Q \subset \Pi$ and disjoint nonempty $P, P' \subseteq Q$, and the prefix of a run of $\calF(Q)$ in which every agent of $Q$ takes Silence.  It is a prefix of a correct run, as $\Lambda$ is empty; every one of its transitions is unary, so none is an interaction between $P$ and $P'$; and nothing is enabled at its last configuration, so no extension of it contains one.

For Entrenched Leader, take disjoint nonempty $P, P' \subset \Pi$, an agent $\ell \notin P \cup P'$, and $Q := P \cup P' \cup \{\ell\}$, and the prefix of a run of $\calF(Q)$ in which every agent of $Q$ elects $\ell$.  Each of its transitions changes one local state, so none is an interaction between $P$ and $P'$.  In any extension, Meet and Elect are disabled and every Relay is over $\{a,\ell\}$ for some $a$, so no transition changes the local state of an agent of $P$ and of an agent of $P'$, and none is an interaction between them.  Neither protocol is therefore interactive$'$.
\end{proof}

\section{Additional Global Platforms}\label{section:additional} 
Beyond social networks, the four-class characterisation applies to diverse global platforms.  Formal classification requires a multiagent atomic transactions specification; most practical platforms lack one.  The grassroots platforms cited below have been formally specified and proven grassroots, hence by Theorems~\ref{theorem:interactive-essential} and~\ref{theorem:interleaving-essential} have all agents but one essential.  The other classifications are informal: we anticipate any reasonable specification will confirm them.

\mypara{Currencies and payment systems} Fiat currencies and Central Bank Digital Currencies (CBDCs) are centralised under government control. PayPal and Venmo similarly have single controlling servers. Bitcoin~\cite{nakamoto2008bitcoin} and Ethereum~\cite{buterin2014next,wood2014ethereum} exemplify decentralised systems with bootstrap nodes. Algorand~\cite{gilad2017algorand} is likewise decentralised: nodes join through a published set of relays, and its peer-to-peer mode discovers peers through DNS seeds and a distributed hash table~\cite{algorand2025p2p}. Ripple~\cite{schwartz2014ripple} operates as federated with designated validators. Grassroots coins and bonds~\cite{shapiro2024gc,shapiro2025atomic,shapiro2026bonds} enable communities to bootstrap local economies from trust relationships.

\mypara{``Sharing economy'' platforms} Uber and Airbnb are centralised with corporate control. OpenBazaar~\cite{openbazaar2014} attempted decentralised peer-to-peer commerce using blockchain but ceased operations. Platform cooperatives~\cite{scholz2016platform} propose federated models where worker-owned nodes coordinate services. Grassroots digital cooperatives~\cite{shapiro2024gc} would enable direct peer-to-peer resource sharing---transportation, housing, labour.

\mypara{Content distribution} YouTube and Netflix are centralised. IPFS~\cite{benet2014ipfs} uses decentralised DHT with bootstrap nodes. PeerTube~\cite{peertube2018} federates video hosting across servers. BitTorrent~\cite{cohen2003incentives} is decentralised: its original version relies on a tracker and its trackerless version on DHT boot nodes.

\mypara{Messaging} WhatsApp and Telegram are centralised. Signal is centralised; its sealed-sender mechanism~\cite{lund2018sealed} is a privacy feature within the centralised architecture. Matrix~\cite{matrix2019} and XMPP~\cite{saint2004rfc} federate across home servers. Scuttlebutt~\cite{tarr2019secure} and Briar~\cite{briar2018} are grassroots with peer-to-peer messaging.

\mypara{Computation} Cloud providers (AWS, Azure) are centralised. Ethereum provides decentralised computation~\cite{buterin2014next,wood2014ethereum}. Grid computing~\cite{foster2001anatomy} federates across institutions. Grassroots Logic Programs~\cite{shapiro2025glp} is a concurrent logic programming language whose transactions-based semantics yields a grassroots protocol.

\mypara{Governance} Existing voting platforms are typically centralised (Change.org). DAOs use decentralised blockchain voting~\cite{wang2019dao}. Grassroots democratic federations~\cite{shapiro2025GF,shapiro2025atomic} support constitutional governance through community formation, sortition-based assembly selection, and multi-level federation.

\mypara{AI and machine learning} OpenAI, Google AI, and Anthropic operate centralised platforms with models hosted exclusively in provider-controlled datacenters. Bittensor~\cite{bittensor2024} achieves decentralisation through blockchain-based coordination with validators and miners. Petals~\cite{borzunov2023petals} approaches grassroots with peer-to-peer inference where volunteers host model shards without central coordination.

\end{document}